\documentclass[preprint,onecolumn,nofootinbib]{revtex4}
\pdfoutput=1
\usepackage[colorlinks=true,linkcolor=blue,urlcolor=blue,filecolor=black,citecolor=red,pdfstartview=FitV,pdftitle={},pdfsubject={},pdfkeywords={},pdfpagemode=None,bookmarksopen=true]{hyperref}
\usepackage{graphicx}
\usepackage{amsmath}
\usepackage{amsfonts}
\usepackage{amssymb,ulem}
\usepackage{color,xcolor}%
\usepackage{subfigure}

\usepackage{dcolumn}
\newcommand{\f}{\begin{equation}}
	\newcommand{\ff}{\end{equation}}
\newcommand{\fa}{\begin{eqnarray}}
	\newcommand{\ffa}{\end{eqnarray}}
\newcommand{\bsub}{\begin{subequations}}
	\newcommand{\esub}{\end{subequations}}

\newcommand{\kk}{\rangle}
\newcommand{\bb}{\langle}
\begin{document}
	
\title{Transport properties of a holographic model with novel gauge-axion coupling}
	\author{Lin-Yue Bai$^{1}$}
	\thanks{bailinyue137@163.com}
	\author{Jian-Pin Wu$^{2}$}
	\thanks{jianpinwu@yzu.edu.cn, corresponding author}
	\author{Zhen-Hua Zhou$^{1}$}
	\thanks{dtplanck@163.com}
	\affiliation{
		$^1$ School of Physics and Electronic Information,Yunnan Normal University, Kunming 650500, China \\
		$^2$ Center for Gravitation and Cosmology, College of Physical Science and Technology, Yangzhou University, Yangzhou 225009, China 
	}	
\begin{abstract}
	
	We investigate the transport properties within a holographic model characterized by a novel gauge-axion coupling. A key innovation is the introduction of the direct coupling between axion fields, the antisymmetric tensor, and the gauge field in our bulk theory. This novel coupling term leads to the emergence of nondiagonal components in the conductivity tensor. An important characteristic is that the off-diagonal elements manifest antisymmetry. Remarkably, the conductivity behavior in this model akin to that of Hall conductivity. Additionally, this model can also achieve metal-insulator transition.
	
\end{abstract}
	\maketitle
	
\section{Introduction}
Gauge/gravity duality, which stands by now as one of the most powerful tools in theoretical physics, provides a deep and fundamental connection between quantum field theory (QFT) and gravity \cite{Maldacena:1997re,Gubser:1998bc,Witten:1998qj,Aharony:1999ti}. This correspondence has revealed some universal properties in strongly coupled quantum many-body systems and has provided important insights into phenomena such as transport properties without quasiparticle excitations, novel mechanisms for superconductivity, and quantum phase transitions (QPTs) \cite{Hartnoll:2009sz,Hartnoll:2016apf}.

Transport is an inherent characteristic of a system, capable of revealing its universal behavior. Extensive study on transport properties has been conducted, as comprehensively reviewed in \cite{Hartnoll:2009sz,Natsuume:2014sfa,Hartnoll:2016apf,Baggioli:2019rrs,Baggioli:2021xuv}. The Kovtun-Son-Starinets density (KSS) is the most successful example, representing the gradient between the shear viscosity and entropy density ratios \cite{Kovtun:2004de, Son:2007vk}. The KSS density has attracted significant attention because it establishes a lower bound for the ratio of shear viscosity to entropy density in any system. This lower bound, known as the KSS bound, provides valuable insights into the behavior of strongly interacting systems and has led to advancements in our understanding of phenomena like the quark-gluon plasma produced in heavy-ion collisions.

To model more realistic systems, we usually need to introduce the momentum dissipation, which removes the $\delta$-function that appears at zero frequency in the electric conductivity of translation-invariant holographic systems. A simple yet significant mechanism incorporating the momentum dissipation is to introduce spatially linear dependent scalar fields, known as axionic fields \cite{Andrade:2013gsa}. The holographic axions model has been extensively applied to address various topics, including the behavior of strange metals \cite{Hartnoll:2009ns,Davison:2013txa,Zhou:2015dha} and the mechanisms underlying coherent and incoherent metals \cite{Davison:2015bea,Zhou:2015qui}. However, it is important to note that when measuring in terms of the chemical potential of the dual field theory, i.e., the system is in the grand canonical ensemble, the direct current (dc) conductivity of the $4$-dimensional holographic system remains a nonvanishing constant independent of temperature \cite{Andrade:2013gsa}. Additionally, it is worth mentioning that a lower bound for the dc conductivity exists in this holographic dual system \cite{Grozdanov:2015qia}. Consequently, this simple holographic axion model does not exhibit a metal-insulator transition (MIT).

In the spirit of effective holographic low energy theories, it is interesting and natural to study the effect of higher-derivative terms of axion fields and the gauge-axion coupling \cite{Gouteraux:2016wxj,Baggioli:2016oqk,Baggioli:2016pia,Li:2018vrz,Huh:2021ppg,Liu:2022bam,Liu:2022bdu,Wang:2021jfu}. Recent studies have revealed that the inclusion of higher-derivative terms in the holographic effective theory can violate the lower bound of dc conductivity in the conventional axion model \cite{Andrade:2013gsa}, providing a more realistic framework to model insulating states with zero dc conductivity at zero temperature.
These higher-derivative terms significantly affect the lower bound of charge diffusion \cite{Baggioli:2016pia,Figueroa:2020tya,Baggioli:2021ejg}, while leaving the upper bound of charge diffusion and energy diffusion unchanged \cite{Huh:2021ppg}. This framework allows for the implementation of a MIT \cite{Baggioli:2016oqk}. Furthermore, two sets of massless axionic fields were considered in this framework \cite{An:2020tkn,Zhou:2018ony}. The first axion field provides a mechanism for momentum dissipation, while the second axion field introduces an elastic mechanical deformation. The electrical conductivity of this case exhibites anomalous inverse Hall conductivity \cite{An:2020tkn,Zhou:2018ony}. Additional discussions of holographic magnetotransport can be referred to \cite{Lucas:2015pxa,Rogatko:2019sjn,Rogatko:2018lhn, Rogatko:2017svr}. In this paper, we propose a novel form of gauge-axion coupling that effectively mimics the effect of a magnetic field. 

The structure for this paper is as follows. In Sec. \ref{2}, we propose a novel holographic effective model, in which a direct coupling between axion fields, the antisymmetric tensor, and the gauge field in introduced. Then the analytical black brane solution is worked out. Moving on to Sec. \ref{3}, we derive the expressions for dc conductivities and engage in a brief discussion of their properties. In Sec. \ref{pro-ec}, we scrutinize the electric conductivities' characteristics, addressing the MIT within both the grand canonical and canonical ensembles. Section \ref{A-magnetic-effect} delves into the exploration of the analogous magnetic effects of our model, accompanied by a comparison with the dyonic model. A comprehensive holographic renormalization procedure is meticulously carried out in appendix \ref{renormalisation}.

\section{The holographic framework}\label{2}

We propose a novel holographic effective model, for which the action is
\begin{align}
	S=\int d^4x\sqrt{-g}\left[R-2\Lambda-\frac{1}{4}F^2-V(X_0)-\frac{\mathcal{J}}{2}Tr[XF]\right]\,.\label{ss}
\end{align}
The above action includes a $U(1)$ gauge field $A_{\mu}$, which is associated with the field strength $F=dA$, and the axion fields $\phi^I$ with $I=1,2$. The axion fields are incorporated into the aforementioned action through the potential $V(X_0)$ and the coupling  $\mathrm{Tr}[XF]$, which are described in the following forms:
\begin{subequations}
	\begin{align}
		&X_0=\frac{1}{2}\delta^{IJ}\partial_\mu\phi^I\partial_\nu\phi^Jg^{\mu\nu}\,,\qquad\qquad\delta^{IJ}=\left(
		\begin{array}{cc}
			~1 ~~&~~ 0~ \\
			~0 ~~&~~ 1~ \\
		\end{array}
		\right) \\
		&\mathrm{Tr}[XF]=\epsilon^{IJ}\partial_\mu\phi^I\partial_\nu\phi^JF^{\mu\nu}\,,\qquad
		\epsilon^{IJ}=\left(
		\begin{array}{cc}
			~0 ~~&~~ 1~ \\
			-1 ~~&~~ 0~ \\
		\end{array}
		\right)
	\end{align}
\end{subequations}
where $\delta^{IJ}$ is the identity matrix and $\epsilon^{IJ}$ is the antisymmetry Levi-Civita symbol. The term $V(X_0)$ exclusively involves the axion fields and has been extensively studied in the holographic effective axions model \cite{Gouteraux:2016wxj,Baggioli:2016oqk,Baggioli:2016pia,Li:2018vrz,Huh:2021ppg,Liu:2022bam,Liu:2022bdu,Wang:2021jfu}. In the effective theory we are investigating, the key innovative aspect is the gauge-axion coupling term, represented as $\mathrm{Tr}[XF]$. This novel coupling term breaks  time reversal invariance, which is discussed in the appendix \ref{renormalisation}.

Usually, the gauge-axion coupling involves the coupling between the axion fields and the ``stress tensor'' of the gauge field $F^2$ through the form as $\mathrm{Tr}[XF^2]=\delta^{IJ}\partial_\mu\phi^I\partial_\nu \phi^J F^{\mu\alpha} F^\nu_{~~\alpha}$, or more direct coupling as $XF^{2}$ \cite{Baggioli:2016pia,Gouteraux:2016wxj}. However, in our model, the axion fields are directly coupled to the gauge field strength $F_{\mu\nu}$. In fact, a more general coupling term, such as $\theta^{IJ}\partial_\mu\phi^I\partial_\nu\phi^JF^{\mu\nu}$ with an arbitrary $2\times 2$ matrix $\theta^{IJ}$, is equivalent for the antisymmetry of $F^{\mu\nu}$.
This novel coupling term produces an equivalent effect to that of a magnetic field, which becomes evident in the subsequent analysis of dc conductivity.

From the above action~\eqref{ss}, the equations of motion can be derived as follows:
\begin{subequations}
	\begin{align}
		\nabla_\mu(V'\partial^\mu\phi^I+\mathcal{J}\epsilon^{IJ}\partial_\nu\phi^JF^{\mu\nu})&=0\,,\\
		\nabla_\mu(F^{\mu\nu}+\mathcal{J}\epsilon^{IJ}\partial^\mu\phi^I\partial^\nu\phi^J)&=0\,,\label{eq2}\\
		R_{\mu\nu}-\frac{1}{2}Rg_{\mu\nu}+\Lambda g_{\mu\nu}&=T_{\mu\nu}\,,
	\end{align}
\end{subequations}
where $V'=dV/dX_0$ and the energy-momentum tensor $T_{\mu\nu}$ reads as:
\begin{align}
	T_{\mu\nu}=&-\frac{1}{2}g_{\mu\nu}(\frac{1}{4}F^2+V(X_0)+\frac{\mathcal{J}}{2}Tr[XF])\nonumber\\
	&+\frac{1}{2}F_{\mu\rho}F^{~~\rho}_\nu +\frac{1}{2}V'(X)\partial_\mu\phi^I\partial_\nu\phi_I+\frac{\mathcal{J}}{2}(\epsilon^{IJ}\partial_\beta\phi^I\partial_\nu\phi^JF_{~~\mu}^\beta+\epsilon^{IJ}\partial_\beta\phi^I\partial_\mu\phi^JF_{~~\nu}^\beta)\,.
\end{align}

The model admits asymptotically AdS charged black brane solutions with the cosmological constant $\Lambda=-3$. For arbitrary choice of the potential function $V(X_0)$, they take the form:
\begin{subequations}
	\begin{align}
		&ds^2=\frac{1}{u^2}(-f(u)dt^2+\frac{1}{f(u)}du^2+dx^2+dy^2)\,,\qquad A=A_t(u)dt+Bxdy\,, \\
		&f(u)=u^3 \int _1^u\frac{1}{4} \left(B^2+2 \mathcal{J} \alpha ^2 B+q^2+\frac{2 V\left(\alpha ^2 \xi^2\right)}{\xi^4}-\frac{12}{\xi^4}\right)d\xi\,,\label{5b}\\
		&A_t(u)=\mu-qu\,,\qquad  \phi^1=\alpha x\,,\qquad \phi^2=\alpha y\,,
	\end{align}	
\end{subequations}
where $u_h$ denotes the horizon location. $\mu$ and $q$ are the chemical potential and the charge density, respectively.
$\alpha$ stands for the strength of the momentum dissipation. The temperature of the black brane is given by

\begin{align}
	&T=-\frac{f'(u_h)}{4\pi}=\frac{3}{4 \pi u_h}-\frac{q^2 u_h^3}{16 \pi }-\frac{V(\alpha ^2 u_h^2)}{8 \pi  u_h}-\frac{\alpha ^2 B J u_h^3}{8 \pi }-\frac{B^2 u_h^3}{16 \pi }\,,\label{T1} 
\end{align}	

Throughout this paper, we will choose $V(X_0)=X_0$. In the equilibrium state, leveraging the renormalization outcomes, we can derive the energy density $\varepsilon =\bb T^{tt}\kk=-2f_3$ and mechanical pressure $\mathcal{P}=\bb T^{xx}\kk=-f_3$. Furthermore, expressions for the entropy density $s$, temperature $T$, energy density $\varepsilon$, and charge density $q$ of the system can be formulated as:
\begin{align}
	&s=\frac{4 \pi }{u_h^2},\quad\quad T=\frac{1}{4\pi u_h} \left(3-\frac{\alpha^2u_h^2}{2}-\frac{u_h^2\mu^2+B^2u_h^4+2 B\mathcal{J}u_h^4\alpha^2}{4}\right),\\
	&\varepsilon=\frac{2}{u_h^3}\left(1-\frac{\alpha^2u_H^2}{2}+\frac{\mu^2u_h^2+B^2u_h^4+2B\mathcal{J}\alpha^2u_h^4}{4}\right),\quad\quad q=\frac{\mu}{u_h}.
\end{align}	
Therefore, the thermodynamic pressure $P$ can be determined using the relation $\varepsilon+P=Ts+q\mu$ as follows:
\begin{align}
 P=\frac{1}{u_h^3}\left(1+\frac{\alpha^2 u_h^2}{2}+\frac{\mu^2 u_h^2-3 B u_h^4-6 B \mathcal{J} \alpha^2 u_h^4}{4}\right).
\end{align}

Then, it is straightforward to obtain the expression of the thermodynamic potential $F(T,V,\mu,B)$ using the relation $F(T,V,\mu,B)=-PV$, where $V=\int dxdy$ denotes the space volume. It is worth noting that the thermodynamic potential can also be worked out through the relation $F(T,V,\mu,B)=-T\ln Z=-TS_E$, where $S_E$ represents the renormalized Euclidean on-shell action. Furthermore, we can validate that these thermodynamic quantities adhere to the following thermodynamic laws:
\begin{align}
	-\frac{\partial F}{\partial V}=P,\quad\quad -\frac{1}{V}\frac{\partial F}{\partial T}=s,\quad\quad -\frac{1}{V}\frac{\partial F}{\partial \mu}=q.
\end{align}   

Utilizing the expression of the thermodynamic potential $F$, the magnetization density $m$ can be computed as follows:
\begin{align}
m=-\frac{1}{V}\frac{\partial F}{\partial B}=-(B+\mathcal{J}\alpha^2)u_h.
\end{align}
When the coupling term is absent, i.e., $\mathcal{J}=0$, the magnetization density $m$ reduces the case of the Einstein-Maxwell theory as elucidated in \cite{Hartnoll:2009sz}. Notably, even in the absence of a magnetic field, the magnetization density $m_0=-\mathcal{J}\alpha^2 u_h$ persists within the system.

In addition, we would like to address the disparity between mechanical pressure and thermodynamic pressure, which is:
\begin{align}
P-\mathcal{P}=\frac{1}{u_h^3}\left(\alpha^2 u_h^2-B^2 u_h^4-2B\mathcal{J} \alpha^2 u_h^4\right).
\end{align}
The difference encompasses the contributions from the magnetic field, the axion field and the coupling term $\mathcal{J}$. When the coupling term is absent, the difference has been discussed in \cite{Hartnoll:2009sz,Ammon:2019apj}.

In the absence of an external magnetic field $B$, it is evident from the expression \eqref{5b} that the novel coupling term $\displaystyle\mathrm{Tr}[XF]$ has no effect on the background. Consequently, the solution described above reduces to a simple Reissner-Nordstrom-AdS (RN-AdS) black brane with axions \cite{Andrade:2013gsa}.
However, it is important to note that this coupling can significantly change the transport behaviors, leading to an off-diagonal electrical conductivity. To better discern the impact of the coupling term on the system, we will consistently deactivate the magnetic field $B$.

\section{DC conductivities}\label{3}

In this section, we will employ the standard membrane technique \cite{Donos:2014uba,Donos:2014cya,Blake:2014yla} to calculate the dc conductivities of the dual field theory. To this end, we turn on the following consistent perturbations around the background:
\begin{align}
	&\delta A_i=(-E_i+A_t(u)\zeta_i)t+\delta a_i(u)~,\nonumber\\
	&\delta g_{ti}=\frac{1}{u^2}(-\zeta_if(u)t+h_{ti}(u))~,\nonumber\\
	&\delta g_{ui}=\frac{h_{ui}(u)}{u^2}~,\nonumber\\
	&\delta\phi_i=\delta\phi_i(u)\,,
\end{align}
where the index $i=x,\,y$ labels the spatial directions. $E_i$ represents the external electric field, while $\zeta_i\equiv-\nabla_iT/T$ denotes the thermal gradient. We would like to emphasize that, because of the novel coupling term, there are always both electric and heat currents along the $y$ direction, with nonzero values. Therefore, for consistency, perturbations along the $y$ direction must be activated to work out the conductivity matrix.

We can use the generalized Ohm's law to calculate the corresponding conductivity coefficients:
\begin{align}
	\left(\begin{array}{cc}
		J &\\
		Q &\\
	\end{array}\right)=\left(\begin{array}{cc}
		\sigma &  \alpha T \\
		\bar{\alpha} T & \bar{\kappa} T \\
	\end{array}\right)\left(\begin{array}{cc}
		E &\\
		-(\nabla T)/T&\\
	\end{array}\right)\,,
\end{align}
where $J$ and $Q$ are electric and heat currents, respectively. $\sigma$ and $\bar{\kappa}$ represent the electric conductivity and the thermal conductivity, respectively. On the other hand, $\alpha$ and $\bar{\alpha}$ denote the thermoelectric conductivity and its reciprocal, where $\alpha=\bar{\alpha}$. Since the bulk currents are conserved, we can evaluated them at the horizon. Consequently, we can obtain the following expressions for the conductivities:
\begin{eqnarray}
	&&
	\sigma_{xx}=\sigma_{yy}=\frac{(V'-\mathcal{J}^2\alpha^2u_h^2)(\mu^2+\alpha^2V')}{\alpha^2(\mathcal{J}^2\mu^2u_h^2+V'^2)}\label{sigmaxx}\,,
	\
	\\
	&&
	\sigma_{xy}=-\sigma_{yx}=\frac{\mathcal{J}\mu u_h(\mu^2+2\alpha^2V'-\mathcal{J}^2\alpha^4u_h^2)}{\alpha^2(\mathcal{J}^2\mu^2u_h^2+V'^2)}\label{sigmaxy}\,,
	\
	\\
	&&
	\alpha_{xx}=\alpha_{yy}=\frac{4\pi\mu(V'-\alpha ^2 \mathcal{J}^2 u_h^2)}{\alpha ^2u_h (\mathcal{J}^2 \mu^2 u_h^2+V'^2)}\,,
	\label{alphaxxyy}
	\
	\\
	&&
	\alpha_{xy}=-\alpha_{yx}=\frac{4\pi\mathcal{J} (\mu^2+\alpha ^2V')}{\alpha ^2 (\mathcal{J}^2 \mu^2 u_h^2+V'^2)}\,,
	\label{alphaxy}
	\
	\\
	&&
	\bar{\kappa}_{xx}=\bar{\kappa}_{yy}=\frac{(4\pi)^2TV'}{\alpha ^2 u_h^2 (\mathcal{J}^2 \mu^2 u_h^2+V'^2)}\,,
	\label{kappaxxyy}
	\
	\\
	&&
	\bar{\kappa}_{xy}=-\bar{\kappa}_{yx}=\frac{(4\pi)^2T\mathcal{J} \mu }{\alpha ^2u_h (\mathcal{J}^2 \mu^2 u_h^2+V'^2)}\,,
	\label{kappaxy}
\end{eqnarray}
where $V'=V'(\alpha^2u_h^2)$. 

An intriguing characteristic we observe is their off-diagonal nature, displaying an antisymmetric conductivity behavior, specifically $\sigma_{xy}=-\sigma_{yx}$. As we know, there are two main mechanisms that can lead to a nonvanished off-diagonal conductivities. One comes from an external magnetic field, the other arises from the anisotropy of the dual system (see, for instance, \cite{An:2020tkn,Ji:2022ovs,Liu:2021stu,Blake:2014yla}). However, in the latter case, the conductivity matrix is symmetric, with $\sigma_{xy}=\sigma_{yx}$ \cite{Ji:2022ovs}, which differs from the earlier mentioned results. On the contrary, the Hall conductivities adhere to the antisymmetric relation $\sigma_{xy}=-\sigma_{yx}$. Hence, we can interpret our findings as ``internal''  Hall conductivities, with the coupled axion fields acting as an ``induced'' magnetic field. There is also the possibility of having an anomalous conductivity, which in the holographic model can be realized by adding a $F\wedge F$ term to the gravity action \cite{Yee:2009vw,Qi:2010qag,Jokela:2012vn,DHoker:2012rlj,Donos:2012js,Seo:2015pug,Seo:2017oyh}.

\section{The properties of the electric conductivities}\label{pro-ec}

In this section, we will delve into the properties of the electric conductivities in our study. Throughout this paper, we will choose $V(X_0)=X_0$. Then the longitudinal conductivity, $\sigma_{xx}$, takes on the following specific form:
\begin{subequations}\label{xy}
	\begin{align}
\sigma_{xx}&=\frac{(\alpha^2+\mu ^2)(1-\mathcal{J}^2 \alpha^2 u_h^2)}{\alpha^2 (\mathcal{J}^2 \mu ^2 u_h^2+1)}\,,\label{sigmaxx-mu}
\
\\
&=\frac{(\alpha^2+q^2u_h^2)(1-\mathcal{J}^2 \alpha^2 u_h^2)}{\alpha^2 (\mathcal{J}^2 q ^2 u_h^4+1)}\,,\label{sigmaxx-q}
\end{align}	
\end{subequations}
and the Hall conductivity, $\sigma_{xy}$, has the form:
\begin{subequations}
	\begin{align}
\sigma_{xy}&=\frac{\mathcal{J} \mu  u_h (\mu ^2+2 \alpha ^2 -\mathcal{J}^2 \alpha ^4 u_h^2)}{\alpha ^2 (\mathcal{J}^2 \mu ^2  u_h^2+1)}\,,\label{sigmaxy-mu}
		\
		\\
	&=\frac{\mathcal{J} q  u_h^2 (q ^2u_h^2+2 \alpha ^2 -\mathcal{J}^2 \alpha ^4 u_h^2)}{\alpha ^2 (\mathcal{J}^2 q ^2  u_h^4+1)}\,.\label{sigmaxy-q}
	\end{align}	
\end{subequations}
Notice that Eqs. \eqref{sigmaxx-mu} and \eqref{sigmaxy-mu} are formulated in terms of the chemical potential $\mu$, while Eqs. \eqref{sigmaxx-q} and \eqref{sigmaxy-q} are expressed in terms of the charge density $q$. We will now analyze two cases of the holographic dual system: one with a fixed chemical potential, corresponding to the grand canonical ensemble, and the other with a fixed charge density, corresponding to the canonical ensemble.

\subsection{Grand canonical ensemble}\label{sec-gce}

In this subsection, we will investigate the transport properties of this holographic system at the grand canonical ensemble, i.e., for a fixed chemical potential $\mu$. When we work in the grand canonical ensemble, we can adopt the chemical potential as the scaling unit, setting it to $\mu=1$. Consequently, Eqs. \eqref{sigmaxx-mu} and \eqref{sigmaxy-mu} become
\begin{subequations}
\begin{align}
\sigma_{xx}&=\frac{(\alpha^2+1)(1-\mathcal{J}^2 \alpha^2 u_h^2)}{\alpha^2 (\mathcal{J}^2  u_h^2+1)}\,,
\label{sigmaxx-mu-1}
\
\\
\sigma_{xy}&=\frac{\mathcal{J}  u_h (1+2 \alpha ^2 -\mathcal{J}^2 \alpha ^4 u_h^2)}{\alpha ^2 (\mathcal{J}^2  u_h^2+1)}\,.
\label{sigma-mu-1}
\end{align}	
\end{subequations}

From Eq. \eqref{sigmaxx-mu-1}, it becomes evident that in order to prevent negative longitudinal conductivity, a constraint of $1 - \mathcal{J}^2 \alpha^2 u_h^2$ should be imposed when the horizon $u_h$ attains its maximum value, i.e., $u_h^2(T=0) = 12/(2 \alpha ^2 + 1)$. This constraint leads to a bound on the parameter $\mathcal{J}$ as follows:
\begin{align}
	\mathcal{J}^2\leq\frac{1}{6}+\frac{1}{12\alpha^2}\,.
	\label{constraint-mu-v1}
\end{align}
It is easy to find that there is a most stringent constraint applicable to all values of $\alpha$ given by:
\begin{eqnarray}
\mathcal{J}^2 \leq 1/6\,.
\label{constraint-mu-v2}
\end{eqnarray}

We would like to point out that the aforementioned constraint \eqref{constraint-mu-v1} also ensures the positivity of $\sigma_{xy}/\mathcal{J}$. This implies that for $\mathcal{J}>0$, an electric field along the positive $x$-axis will induce a current in the positive $y$-axis direction. Conversely, for $\mathcal{J}<0$, an electric field along the positive $x$-axis will result in a current opposite to the direction of the $y$-axis, which is similar to the effects of Lorentz forces.

\begin{figure}[ht]
	\centering
	\includegraphics[width=8cm]{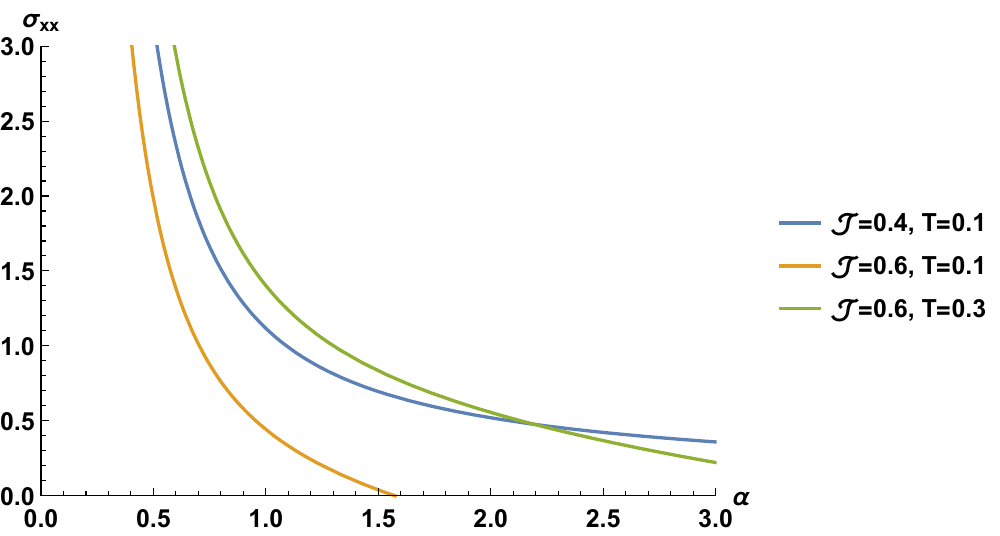}\quad  \includegraphics[width=8cm]{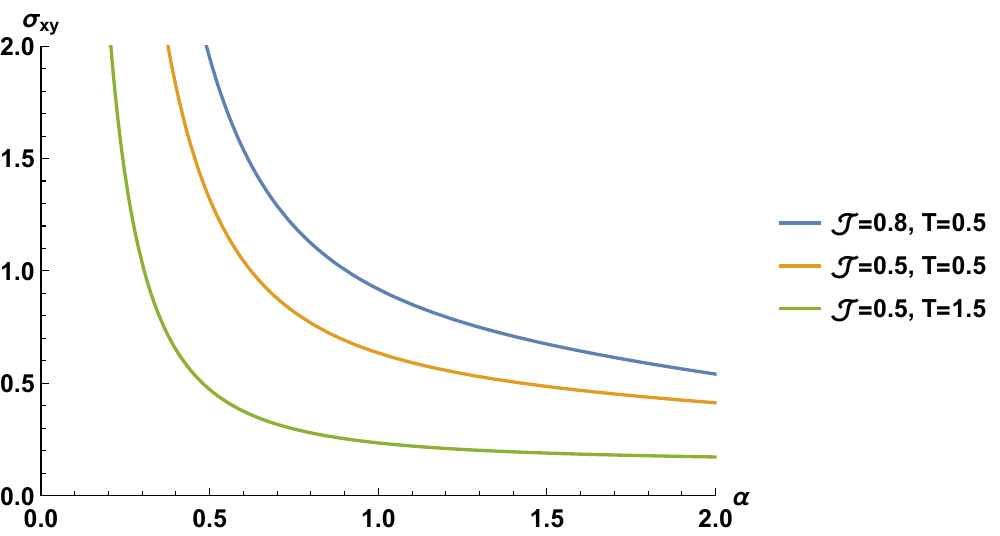} \\
	\caption{The longitudinal conductivity, $\sigma_{xx}$, and the Hall conductivity, $\sigma_{xy}$, as functions of the disorder strength $\alpha$ in the grand canonical ensemble. Here we have keep the coupling parameter $\mathcal{J}$ and the temperature fixed.}\label{ddvsk}
\end{figure}

Subsequently, we present the dependence of the longitudinal conductivity, $\sigma_{xx}$, and the Hall conductivity, $\sigma_{xy}$, on the disorder strength $\alpha$ while keeping the coupling parameter $\mathcal{J}$ and temperature fixed in Fig.\ref{ddvsk}. Our findings indicate a decreasing trend in both the longitudinal and Hall conductivities with increasing $\alpha$, signifying the suppression of conductivity by disorder. This observation is consistent with conventional axion models such as those found in \cite{Andrade:2013gsa,Gouteraux:2016wxj,Baggioli:2016oqk,Baggioli:2016pia,Li:2018vrz,Huh:2021ppg,Liu:2022bam,Liu:2022bdu,Wang:2021jfu}.

\begin{figure}[ht]
	\centering
\includegraphics[width=8cm]{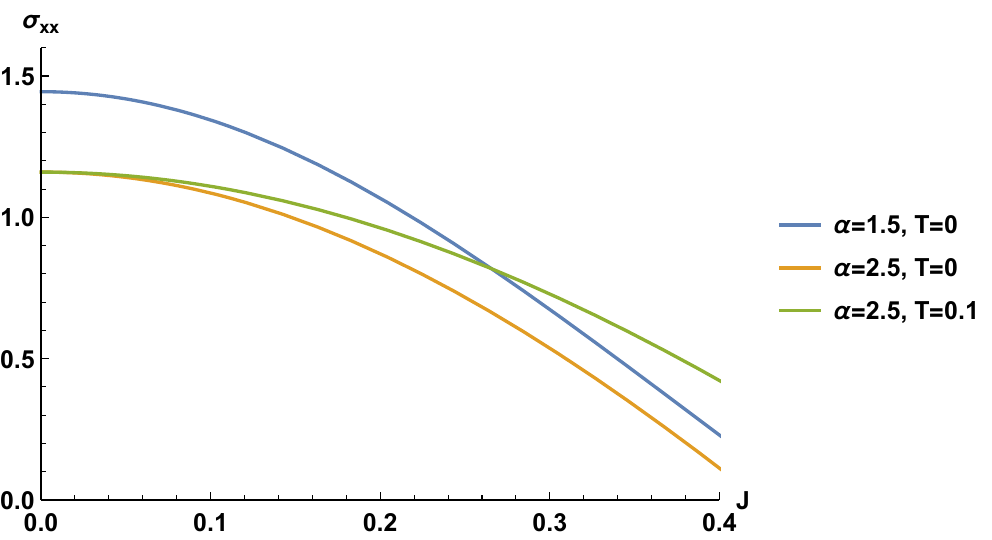}\quad  \includegraphics[width=8cm]{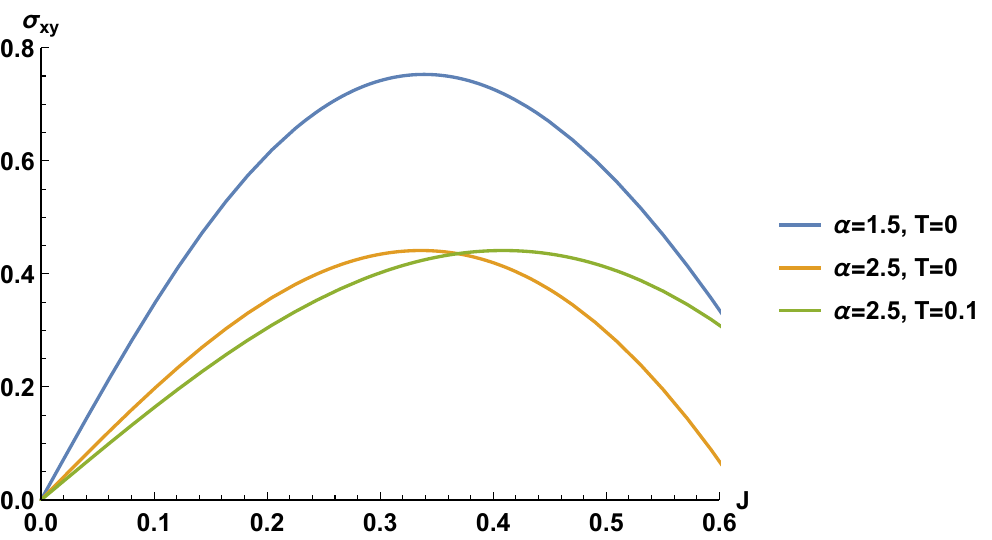} \\
	\caption{The longitudinal conductivity, $\sigma_{xx}$, and the Hall conductivity, $\sigma_{xy}$, as functions of the coupling strength $\mathcal{J}$ in the grand canonical ensemble. Here we have keep the disorder strength $\alpha$ and the temperature fixed.
	}\label{ddvsJ}
\end{figure}

Furthermore, we depict the relationship between the longitudinal conductivity and the Hall conductivity with respect to the coupling parameter $\mathcal{J}$ while keeping the disorder strength $\alpha$ and temperature fixed (Fig. \ref{ddvsJ}). Our observations reveal that within the regime of small $\mathcal{J}$ values, the longitudinal conductivity decreases, whereas the Hall conductivity increases as $\mathcal{J}$ is raised. During this phase, the coupling term $\mathcal{J}$ plays a role akin to the Lorentz force induced by a magnetic field. However, as $\mathcal{J}$ enters the large regime, both the longitudinal and Hall conductivities are suppressed by the increasing coupling strength $\mathcal{J}$. This behavior may resemble the dominant onset of the vortex magnetoresistance effect.

\begin{figure}[tbh]
	\center{
		\includegraphics[width=8cm]{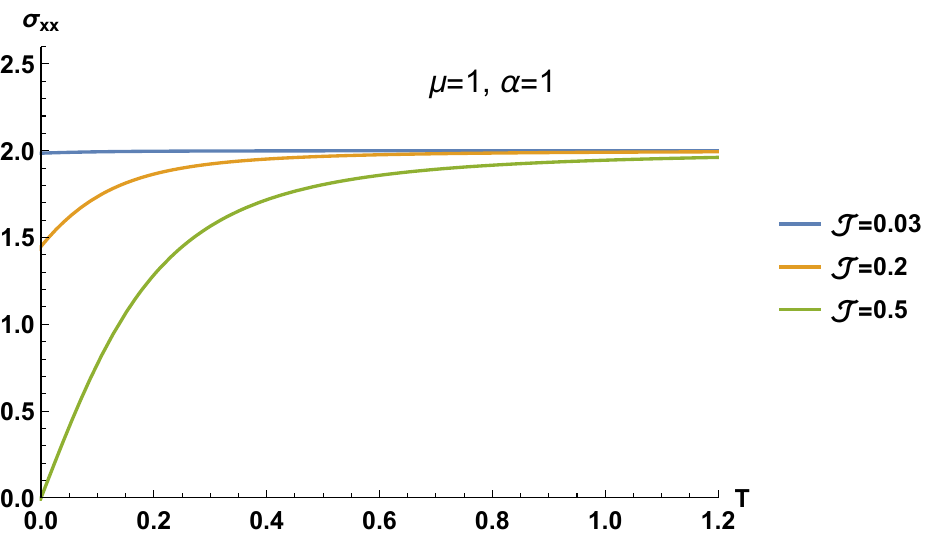}\quad
		\includegraphics[width=8cm]{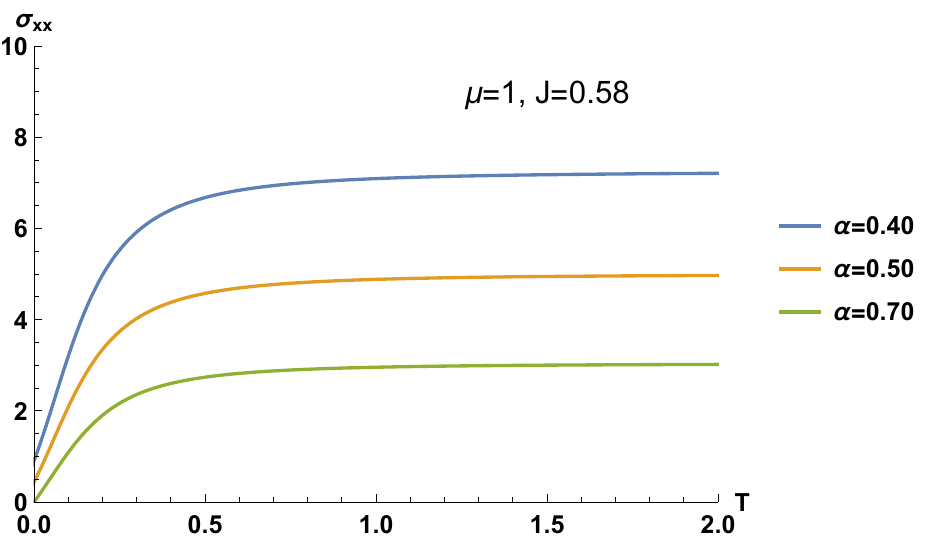} 
		\caption{\label{XFMIT}The temperature behaviors of $\sigma_{xx}$ for various $\alpha$ and $\mathcal{J}$.}}
\end{figure}

Now, we would like to study the transport behavior of this system. To identify the metallic and insulating phases, we adopt a widely used operational definition, as described in several holographic references \cite{Baggioli:2016rdj,Donos:2012js,Donos:2013eha,Donos:2014uba,Ling:2015ghh,Ling:2015dma,Ling:2015epa,Ling:2015exa,Ling:2016wyr,Ling:2016dck,Baggioli:2014roa,Baggioli:2016oqk,Baggioli:2016oju,Donos:2014oha,Liu:2021stu,Kiritsis:2015oxa,Liu:2022bdu,Li:2022yad}. Specifically, $\partial_{T}\sigma_{xx}<0$ will indicate the metallic phase, while $\partial_{T}\sigma_{xx}>0$ will indicate the insulating phase. Figure \ref{XFMIT} illustrates the temperature dependence of $\sigma_{xx}$ for different values of $\mathcal{J}$ and $\alpha$. It is evident that as the temperature decreases, $\sigma_{xx}$ also decreases. This observation suggests that the holographic system under investigation displays an insulating behavior.

We can also analytically calculate the temperature derivative of $\sigma_{xx}$ at finite temperature as
\begin{equation}
	\partial_{T}\sigma_{xx}=\frac{32\pi \mathcal{J}^2 u_h^3(1+\alpha^2)^2}{\alpha^2(12+2\alpha^2u_h^2+u_h^2)(1+\mathcal{J}^2u_h^2)^2}\,.
	\label{sigma-d-tem-mu-1}
\end{equation}
It is evident that $\partial_{T}\sigma_{xx}>0$, indicating that this holographic dual system indeed demonstrates insulating behavior, which aligns with the findings presented in Fig.\ref{XFMIT}. Additionally, we have observed that in the high-temperature limit, $\sigma_{xx}$ approaches a maximum value as
\begin{equation}
	\sigma(T\rightarrow\infty)=1+\frac{1}{\alpha^2}\,.
	\label{sigma-h-tem-mu-1}
\end{equation}
This indicates an upper bound for the conductivity, which is dependent on $\alpha$ but independent of $\mathcal{J}$. This result is in agreement with the behavior depicted in Fig.\ref{XFMIT}.

It is also interesting to investigate the conductive behavior at zero temperature. 
Figure \ref{XFMIT} demonstrates that in certain parameter region, $\sigma_{xx}$ can approach zero in the limit of zero temperature, indicating the emergence of an ideal insulator. Next, we will delve into a comprehensive analysis of the zero-temperature conductivity behavior. To this end, we can analytically work out the expression of the longitudinal conductivity at zero temperature, which takes the form:
\begin{align}
	\sigma_{xx}(T=0)=\frac{(1+\alpha^2)(1+2\alpha^2(1-6\mathcal{J}^2))}{2\alpha^4+\alpha^2(1+12\mathcal{J}^2)}\,.\label{ddxxT0}
\end{align}

\begin{figure}[ht]
	\centering
	\includegraphics[width=10cm]{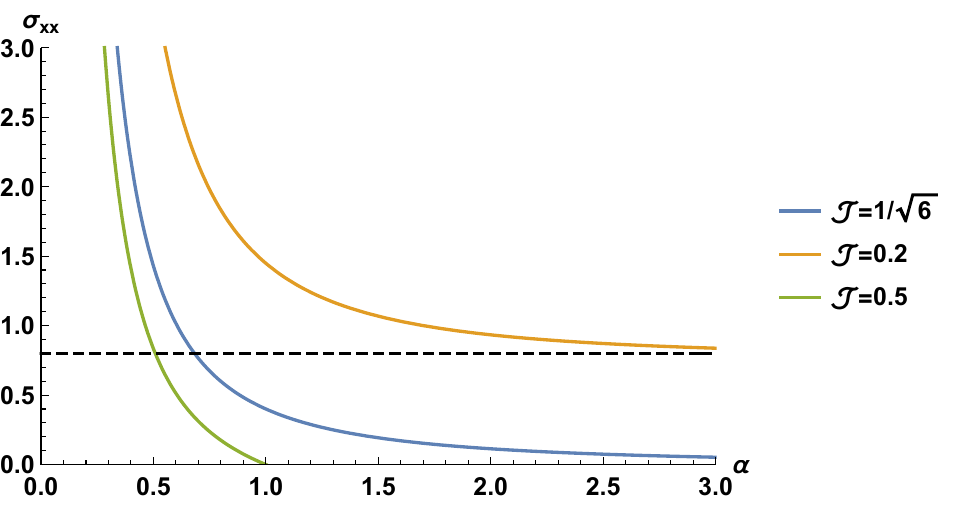}\\
	\caption{$\sigma_{xx}(T=0)$ as a function of $\alpha$ for various coupling parameters $\mathcal{J}$ in the grand canonical ensemble.}\label{ddmiu1T0}
\end{figure}

First, when we take the upper bound of the constraint \eqref{constraint-mu-v2}, i.e, $\mathcal{J}^2=1/6$, it becomes evident from Eq. \eqref{ddxxT0} that $\sigma_{xx}(T=0)$ that $\sigma_{xx}(T=0)$ decreases with increasing disorder strength $\alpha$ and eventually tends to zero in the large $\alpha$ limit, indicating the emergence of an ideal insulator. This process is prominently illustrated in Fig.\ref{ddmiu1T0}. In this case, the realization of the ideal insulator is predominantly driven by the strength of disorder. 

Moreover, for cases where $\mathcal{J}^2<1/6$, the conductivity $\sigma_{xx}(T=0)$ remains positive for all values of $\alpha$, bounded by $\sigma_{xx}(T=0)=1-6\mathcal{J}^2$. Consequently, the dual system is regarded as a poor insulating phase. An illustrative instance of this is presented in Fig. \ref{ddmiu1T0}, where we consider $\mathcal{J}=0.2$. 

In instances where $\mathcal{J}^2>1/6$, $\sigma_{xx}(T=0)$ approaches zero at a finite value of $\alpha$ as shown in Fig.\ref{ddmiu1T0} for $\mathcal{J}=0.5$, implying the potential attainment of an ideal insulator. However, it is important to emphasize that this ideal insulating phase is not primarily driven by disorder but might be associated with some dynamical instabilities. In fact, from Eq.\eqref{ddxxT0}, it becomes evident that when the relation $\mathcal{J}^2=1/6+1/(12\alpha^2)$ is satisfied, $\sigma_{xx}=0$ in the zero temperature limit.

In conclusion, the holographic dual system \eqref{ss} in the grand canonical ensemble demonstrates insulating behavior irrespective of the disorder strength $\alpha$ and the coupling parameter $\mathcal{J}$. However, at zero temperature, the dual system displays either ideal or poor insulating behavior, contingent upon the value of $\mathcal{J}$. Specifically, an ideal insulating phase can be realized either when $\mathcal{J}^2=1/6$ with large $\alpha$ or when $\mathcal{J}^2=1/6+1/(12\alpha^2)$. Conversely, in instances where $\mathcal{J}^2<1/6$, the dual system exhibits poor insulating properties.

\subsection{Canonical ensemble}\label{can-ens}

In this subsection, we turn to the canonical ensemble, within which we are able to set the charge density as $q=1$. Consequently, in this scenario, the conductivities can be expressed as follows:
\begin{subequations}
	\begin{align}
		\sigma_{xx}&=\frac{(\alpha^2+u_h^2)(1-\mathcal{J}^2 \alpha^2 u_h^2)}{\alpha^2 (\mathcal{J}^2 u_h^4+1)}\,,
		\label{sigmaxx-q-1}
		\
		\\
		\sigma_{xy}&=\frac{\mathcal{J}   u_h^2 (u_h^2+2 \alpha ^2 -\mathcal{J}^2 \alpha ^4 u_h^2)}{\alpha ^2 (\mathcal{J}^2  u_h^4+1)}\,.
		\label{sigma-q-1}
	\end{align}	
\end{subequations}

To guarantee the non-negativity of $\sigma_{xx}$, even when $u_h^2$ takes its maximum value of $u_h^2=\sqrt{\alpha^4+12}-\alpha^2$, which corresponds to the scenario of zero temperature, the following constraint comes into play:
\begin{align}
	\mathcal{J}^2\leq\frac{1}{12}+\frac{\sqrt{\alpha^4+12}}{12\alpha^2}\,.\label{csJq}
\end{align}
Different from the case of the grand canonical ensemble, under the above constraint, the Hall conductivity $\sigma_{xy}$ is also positive. In addition, similarly to the scenario in the grand canoncical case, we find that there is also a tightest constraint $\mathcal{J}^2\leq 1/6$, which is available for all $\alpha$.
\begin{figure}[ht]
	\centering
	\includegraphics[width=8cm]{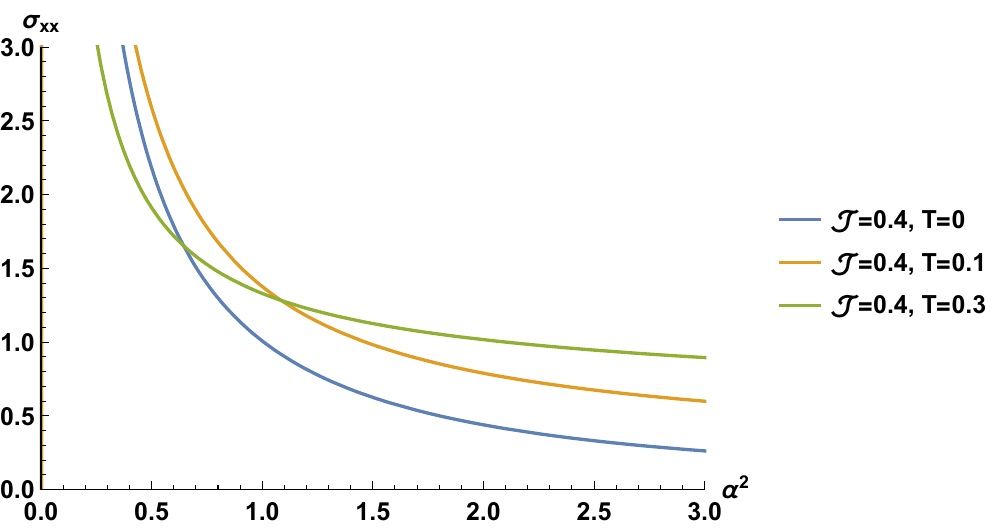}\quad \includegraphics[width=8cm]{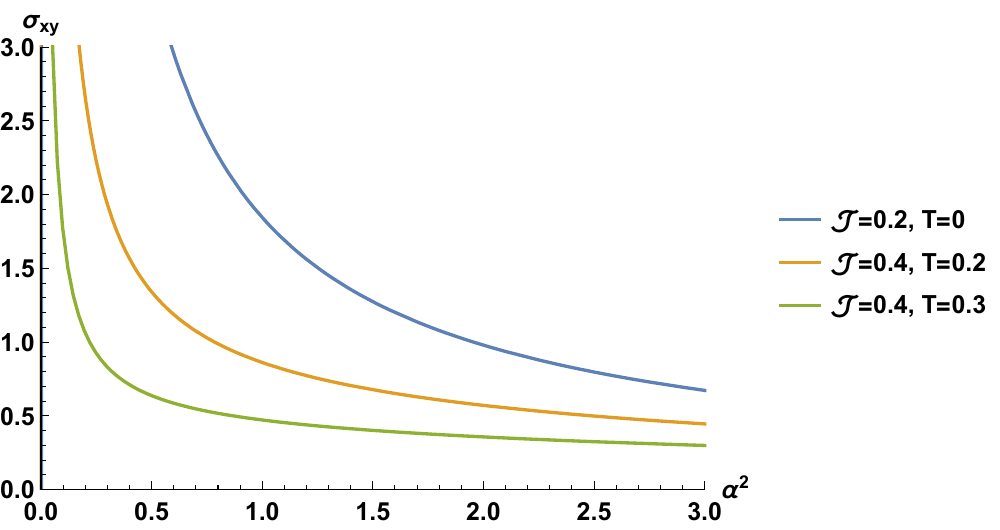} \\
	\caption{The longitudinal conductivity, $\sigma_{xx}$, and the Hall conductivity, $\sigma_{xy}$, as functions of the disorder strength $\alpha$ in the canonical ensemble. Here we have keep the coupling parameter $\mathcal{J}$ and the temperature fixed.}\label{ddqvsk}
\end{figure}

\begin{figure}[ht]
	\centering
	\includegraphics[width=8cm]{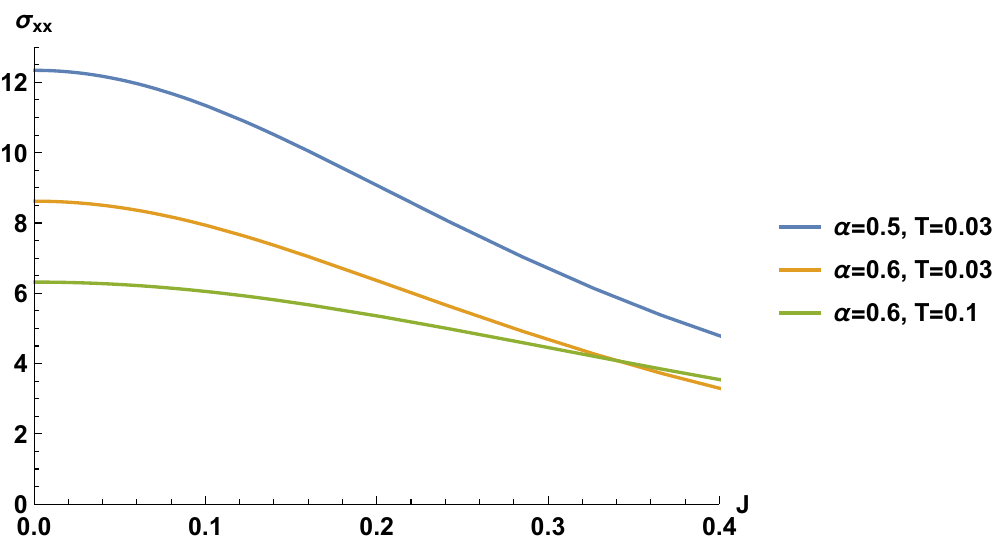}\quad \includegraphics[width=8cm]{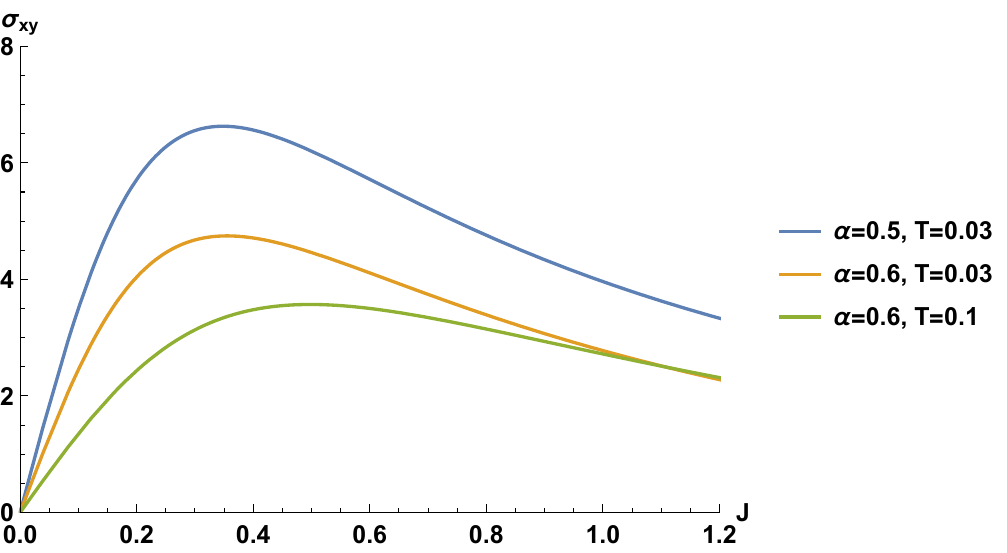}\\
	\caption{The longitudinal conductivity, $\sigma_{xx}$, and the Hall conductivity, $\sigma_{xy}$, as functions of the coupling strength $\mathcal{J}$ in the grand canonical ensemble. Here we have keep the disorder strength $\alpha$ and the temperature fixed.}\label{ddqvsJ}
\end{figure}

We also illustrate how the longitudinal conductivity and the Hall conductivity vary on the disorder strength $\alpha$ or the coupling parameter $\mathcal{J}$ while keeping the other parameters and the temperature fixed, which are showcased in Fig.\ref{ddqvsk} and Fig.\ref{ddqvsJ}. We observe that the dependence of the longitudinal conductivity and the Hall conductivity on the disorder strength $\alpha$ or the coupling parameter $\mathcal{J}$ is similar to the case in the grand canonical ensemble studied in the above subsection.

Next, we will delve into studying the transport properties of this holographic dual system in the canonical ensemble. To this end, we we will derive the expression of $\partial_{T}\sigma_{xx}$ as follows:
\begin{align}
	\partial_{T}\sigma_{xx}=-\frac{32\pi u_h^3(\mathcal{J}^2(\mathcal{J}^2\alpha^4-1)u_h^4-4\mathcal{J}^2\alpha^2u_h^2+1-\mathcal{J}^2\alpha^4)}{\alpha^2(1+\mathcal{J}^2u_h^4)^2(12+3u_h^4+2\alpha^2u_h^2)}\,.
	\label{dsigmadT-q1}
\end{align}
From the equation above, we observe that in order to investigate the temperature-dependent behaviors of $\sigma_{xx}$, our focus should primarily be on analyzing the numerator:
{\begin{align}
Y(u_h^2)=\mathcal{J}^2(1-\mathcal{J}^2\alpha^4)u_h^4+4\mathcal{J}^2\alpha^2u_h^2+\mathcal{J}^2\alpha^4-1\,.
\label{Yuh}
\end{align}
Notice that $Y(u_h^2)$ is a quadratic function of $u_h^2$. We can work out the roots of $Y(u_h^2)=0$, which are given by:
\begin{subequations}
	\begin{align}
		u_h^2&=(1+\mathcal{J}\alpha^2)/(\mathcal{J}^2\alpha^2-\mathcal{J})\,,
		\label{uhv1}
		\
		\\
		u_h^2&=(1-\mathcal{J}\alpha^2)/(\mathcal{J}^2\alpha^2+\mathcal{J})\,.
		\label{uhv2}
	\end{align}	
\end{subequations}
Without loss of generality, we assume $\mathcal{J}\geq0$ and categorize our analysis into two distinct case: $\mathcal{J}\alpha^2<1$ and $\mathcal{J}\alpha^2>1$, for further discussions. We visually represent the behavior of $Y(u_h^2)$ in Fig.\ref{Yvsuh2}.
\begin{figure}[ht]
	\centering
	\includegraphics[width=7cm]{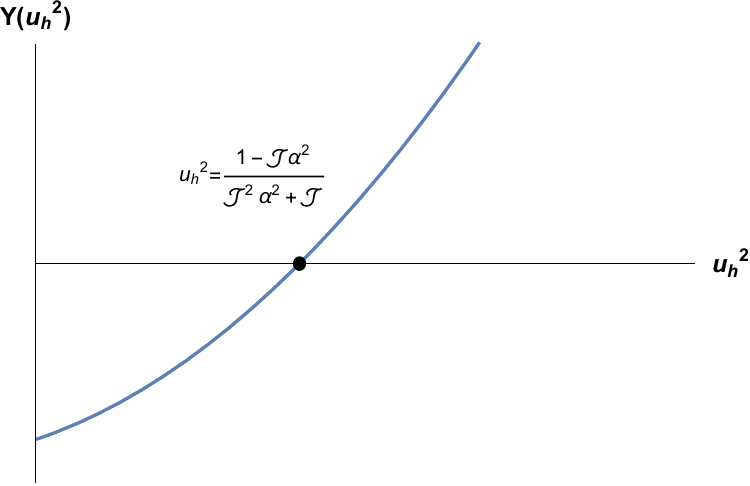}\quad\quad \includegraphics[width=7cm]{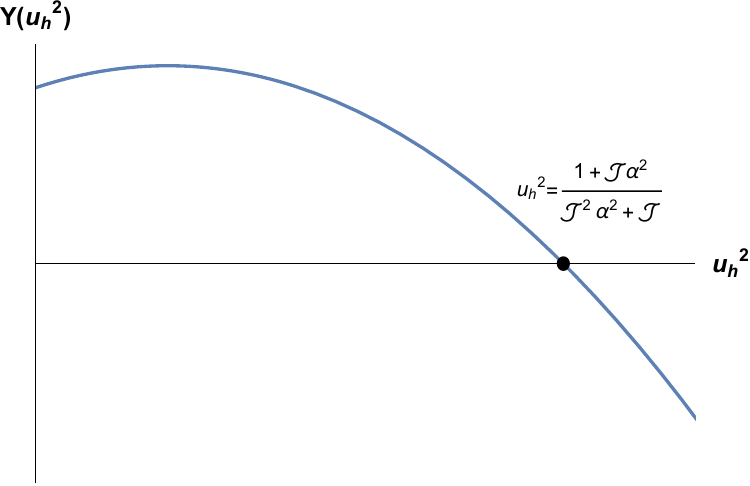}\\
	\caption{$Y(u_h^2)$ varies with respect to $u_h^2$. The left panel corresponds to $\mathcal{J}\alpha^2<1$, while the right panel is for $\mathcal{J}\alpha^2>1$.}\label{Yvsuh2}
\end{figure}

Before we proceed, it is important to note that $u_h$ increases from zero, corresponding to the high-temperature case, to a certain finite value of $u_h$, indicating the zero-temperature case. Now, we turn to the analytical study of the conductivity behavior of this system. 
First, when the maximun value of $u^2_h$, which is $u^2_h=\sqrt{\alpha^4+12}-\alpha^2$ at zero temperature, is smaller than the positive root,  $\partial_{T}\sigma_{xx}$ is consistently negative for $\mathcal{J}\alpha^2<1$ or positive for $\mathcal{J}\alpha^2>1$. This indicates that the system exhibits metallic behavior for $\mathcal{J}\alpha^2<1$ and insulating behavior for  $\mathcal{J}\alpha^2>1$. Otherwise, when the maximum value exceeds the positive root, the multiple phases can be observed. We present all possible scenarios in the following list based on different parameter regions:
\begin{align*}
	\mathcal{J}\alpha^2<1\qquad \begin{cases} ~(1)~\dfrac{1-\mathcal{J}\alpha^2}{\mathcal{J}^2\alpha^2+\mathcal{J}}>\sqrt{\alpha^4+12}-\alpha^2 &  \dfrac{d\sigma}{dT}<0  ~~\text{always}\\\\
		~(2)~\dfrac{1-\mathcal{J}\alpha^2}{\mathcal{J}^2\alpha^2+\mathcal{J}}<\sqrt{\alpha^4+12}-\alpha^2 &  \dfrac{d\sigma}{dT}>0\rightarrow \dfrac{d\sigma}{dT}<0 ~~\text{from $T=0$ to high $T$}\end{cases}\\\\
	\mathcal{J}\alpha^2>1\qquad \begin{cases}  ~(3)~\dfrac{1+\mathcal{J}\alpha^2}{\mathcal{J}^2\alpha^2-\mathcal{J}}>\sqrt{\alpha^4+12}-\alpha^2 &  \dfrac{d\sigma}{dT}>0  ~~\text{always}\\\\
		~(4)~\dfrac{1+\mathcal{J}\alpha^2}{\mathcal{J}^2\alpha^2-\mathcal{J}}<\sqrt{\alpha^4+12}-\alpha^2 &  \dfrac{d\sigma}{dT}<0\rightarrow \dfrac{d\sigma}{dT}>0 ~~\text{from $T=0$ to high $T$}\end{cases}
\end{align*}
Case (1) corresponds to a metallic phase, and case (3) corresponds to an insulating phase. In case (2), as the temperature decreases, the system transitions from the metallic phase to the insulating phase. Conversely, in case (4), the electrical conductivity behavior exhibits the opposite trend compared to case (2). In other words, as the temperature decreases in case (4), the system undergoes a transition from the insulating phase to the metallic phase.

However, it is important to note that case (4) does not exist, as demonstrated in the following proof.
From \eqref{csJq} and the condition $ \mathcal{J}\alpha^2>1$, we derive the following inequalities:
\begin{align*}
	\mathcal{J}^2\leq\frac{1}{12}+\frac{1}{12}\sqrt{1+\frac{12}{\alpha^4}}<\frac{1}{12}+\frac{1}{12}\sqrt{1+12\mathcal{J}^2}\,,
\end{align*}
which implies $\mathcal{J}^2<1/4$. On the other hand, we have
\begin{align*}
	&\dfrac{1+\mathcal{J}\alpha^2}{\mathcal{J}^2\alpha^2-\mathcal{J}}=\frac{1}{\mathcal{J}}(1+\frac{2}{\mathcal{J}\alpha^2-1})>\frac{1}{\mathcal{J}}\,,\\
	&\sqrt{\alpha^4+12}-\alpha^2<\sqrt{\frac{1}{\mathcal{J}^2}+12}-\frac{1}{\mathcal{J}}
\end{align*}
and therefore, the condition $1/\mathcal{J}>\sqrt{1/\mathcal{J}^2+12}-1/\mathcal{J}$ is satisfied when $\mathcal{J}^2<1/4$, which establishes the claim. 

For the sake of visualization, we illustrate the first three cases in Fig.\ref{ddqTT}. As shown in Fig.\ref{ddqTT}, it is evident that the ideal insulator state can be achieved in both case (2) and case (3).
\begin{figure}[ht]
	\centering
	\includegraphics[width=5cm]{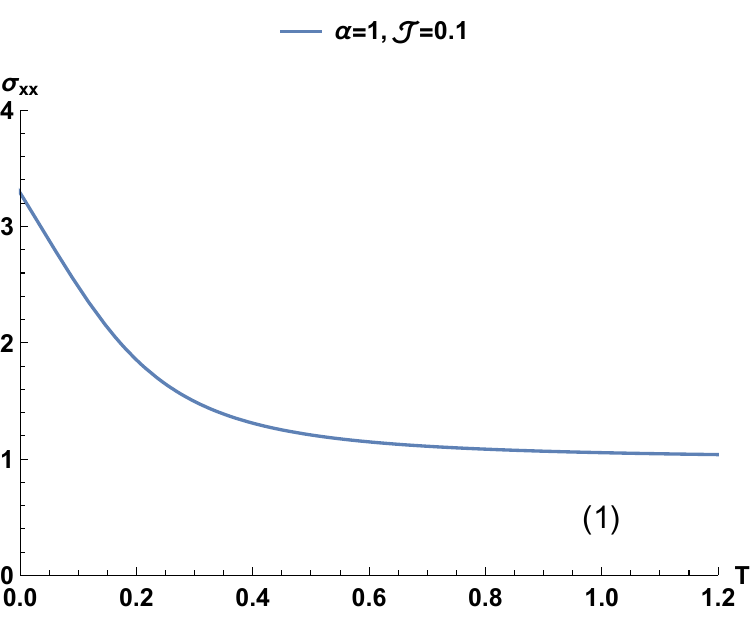}\quad   \includegraphics[width=5cm]{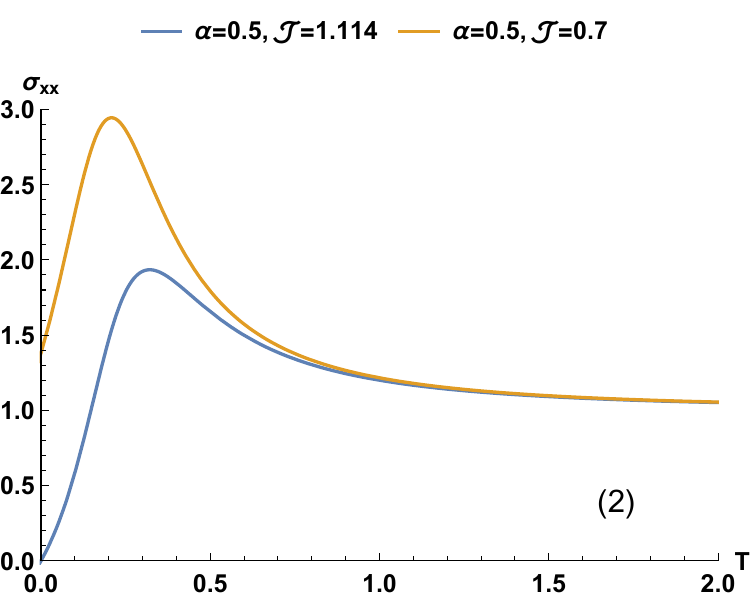} \quad  \includegraphics[width=5cm]{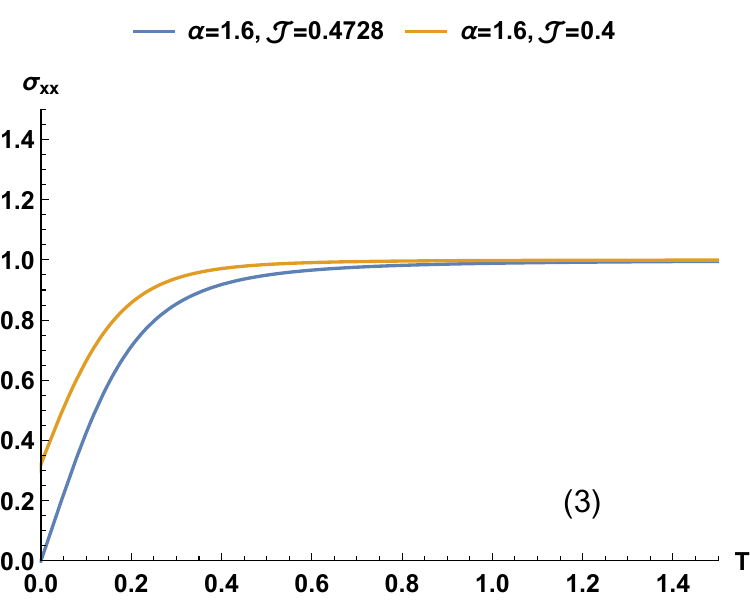}\\
    \caption{$\sigma_{xx}$ as a function of $T$ for different three different cases.}\label{ddqTT}
\end{figure}

\section{Analogous magnetic effect}\label{A-magnetic-effect}

As mentioned in Sec. \ref{3}, it has been noted that the conductivities within the holographic system exhibit off-diagonal behavior, notably $\sigma_{xy}=-\sigma_{yx}$, a phenomenon induced by the novel coupling term $\displaystyle\mathrm{Tr}[XF]$. In this section, we will delve into its analogous magnetic effects and explore the distinctions between the magnetic effects arising from an external magnetic field and those originating from this novel coupling term.

\subsection{Analogous magnetic effect}

Let us consider the following replacement:
\begin{align}
	V(X_0)\rightarrow V(X_0)+\frac{1}{2}\mathcal{J}^2X^2_0\,,\qquad \mathcal{J}\rightarrow\frac{B}{\alpha^2}\,,\label{substi}
\end{align}
where $B$ represents an equivalent magnetic field. Then, the conductivities~\eqref{sigmaxx} and \eqref{sigmaxy} change to
\begin{align}
	\sigma_{xx}&=\sigma_{yy}=\frac{\alpha^2V'(B^2u_h^2+\alpha^2V'+\mu^2)}{(B^2u_h^2+\alpha^2V')^2+B^2u_h^2\mu^2}\,,\label{xxcom}\\
	\sigma_{xy}&=-\sigma_{yx}=\frac{B\mu u_h(B^2u_h^2+2\alpha^2V'+\mu^2)}{(B^2u_h^2+\alpha^2V')^2+B^2u_h^2\mu^2}\,.\label{xycom}
\end{align}
refer to \cite{An:2020tkn}. We observe that the obtained result coincides with that derived from the action \eqref{ss} with $\mathcal{J}=0$, but with an external magnetic field $A=A_tdt+Bxdy$. Therefore, we can interpret $\mathcal{J}\alpha^2$ as an effective magnetic field $B$. As the coupling does not affect the background solution, the effective magnetic field $\mathcal{J}\alpha^2$ is induced by the external electrical perturbation. It may represent some induced magnetic momentum of the dual systems.

The thermal and thermoelectric conductivities also exhibit the similar characteristics to the electric conductivities: the off-diagonal nature with an antisymmetric behavior. When we apply the replacement \eqref{substi}, both the thermal and thermoelectric conductivities transform into those derived from the action \eqref{ss} with $\mathcal{J}=0$, but with the inclusion of an external magnetic field $B$.
\begin{figure}[tbh]
	\center{
		\includegraphics[width=10cm]{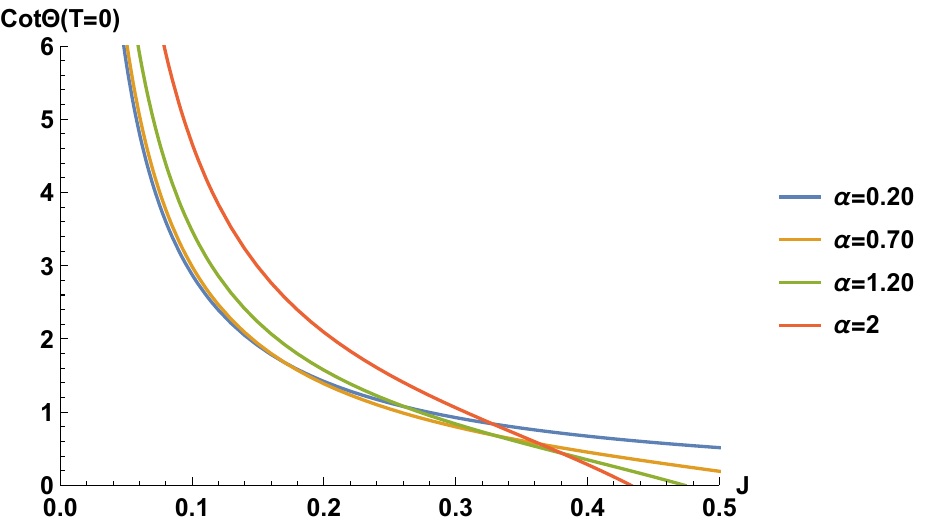}
		\caption{\label{XFB} The inverse Hall angle as a function of coupling parameter $\mathcal{J}$ for different $\alpha$ in the grand canonical ensemble at $T=0$.}}
\end{figure}

Given the presence of an analogous magnetic effect within this dual system, it becomes intriguing to examine the inverse Hall angle, defined as $\cot{\Theta}\equiv\sigma_{xx}/\sigma_{xy}$, which can be used to measure the relative magnitudes of these two conductivities. We focus on the inverse Hall Angle at zero temperature, which can be expressed as follows:
\begin{align}
	\cot{\Theta}=-\frac{\left(\mu ^2+\alpha ^2\right) \sqrt{\mu ^2+2 \alpha ^2} \left( \left(\mu ^2+2 \alpha ^2 \right)-12 \alpha ^2 \mathcal{J}^2\right)}{2 \sqrt{3} \mathcal{J} \mu  \left(12 \alpha ^4 \mathcal{J}^2-\left(\mu ^2+2 \alpha ^2\right)^2\right)}\,.
\end{align}
then we show the inverse Hall angle $\cot{\Theta}$ as a function of the coupling $\mathcal{J}$ in the grand canonical ensemble in Fig.\ref{XFB}\footnote{The behavior of the inverse Hall angle exhibits similarity within the canonical ensemble as well.}. We observe that as $\mathcal{J}$ increases, the inverse Hall angle $\cot{\Theta}$ decrease, while maintaining a fixed disorder strength $\alpha$. However, as $\mathcal{J}$ decreases and approaches zero, we observe a rapid increase in the inverse Hall angle, approaching infinity. This suggests an increasingly pronounced Hall conductivity $\sigma_{xy}$ during the process of $\mathcal{J}$ approaching zero. We would like to emphasize that the observed monotonic decrease of the inverse Hall angle as $\mathcal{J}$ is increased aligns with the behavior observed in the model with a magnetic field \cite{An:2020tkn,Zhou:2018ony}. This provides further confirmation that the novel coupling term indeed induces an effect equivalent to that of a magnetic field.

\subsection{The electrical transports in a simple dyonic model}

In this subsection, we will provide a brief examination of electrical transport behaviors within a simple dyonic model employing conventional axion fields. These behaviors will be studied within the canonical ensemble and then compared to those exhibited by our present model with the novel coupling term. For this simple dyonic model  , which refer to \cite{Donos:2014uba}, the expressions for temperature and conductivities are as follows:
\begin{align}
	&T=\frac{1}{4 \pi u_h}\Big(3-\frac{\alpha^2 u_h^2}{2 }-\frac{(1+B^2) u_h^4}{4 }\Big)\\
	&\sigma_{xx}=\sigma_{yy}=\frac{\alpha^2(B^2u_h^2+\alpha^2+u_h^2)}{(B^2u_h^2+\alpha^2)^2+B^2u_h^4}\\
	&\sigma_{xy}=-\sigma_{yx}=\frac{Bu_h^2(B^2u_h^2+2\alpha^2+u_h^2)}{(B^2u_h^2+\alpha^2)^2+B^2u_h^4}
\end{align}
Without loss of generality, we shall assume that $B\geq 0$ in the following.

When $T=0$, we find that $u_h^2=(\sqrt{\alpha^4+12B^2+12}-\alpha^2)/(1+B^2)$, leading to the following expressions for the conductivities at $T=0$:
\begin{align}
	&\sigma_{xx}(T=0)=\frac{\alpha^2 \sqrt{\alpha^4+12 B^2+12}}{12 B^2+\alpha^4}\\
	&\sigma_{xy}(T=0)=\frac{12 B}{12 B^2+\alpha^4}
\end{align}
It is evident that both $\sigma_{xx}(T=0)$ and $\sigma_{xy}(T=0)$ are strictly greater than zero, and $\sigma_{xx}(T=0)$ exhibits a lower bound of $1$ as $\alpha$ increases.

Additionally, we calculate the derivatives of both conductivities, $\sigma_{xx}(T=0)$ and $\sigma_{xy}(T=0)$, with respect to $\alpha$, resulting in the following expressions:
\begin{align}
	&\frac{d\sigma_{xx}(T=0)}{d\alpha}=\frac{24 \alpha(12 B^4+B^2 (\alpha^4+12)-\alpha^4)}{(12 B^2+\alpha^4)^2 \sqrt{12 B^2+\alpha^4+12}}\\
	&\frac{d\sigma_{xy}(T=0)}{d\alpha}=-\frac{48 B \alpha^3}{(12 B^2+\alpha^4)^2}
\end{align}
From the above equations and Fig.\ref{dyonicddq1T0}, we observe distinct behaviors in $\sigma_{xx}(T=0)$ and $\sigma_{xy}(T=0)$ as $\alpha$ varies. For $B<1$, $\sigma_{xx}(T=0)$ initially increases with rising $\alpha$ and then decreases as $\alpha$ continues to increase. Conversely, when $B\geq 1$, $\sigma_{xx}(T=0)$ consistently and monotonically increases, eventually converging toward an upper limit of $1$ for large values of $\alpha$. These patterns are visually illustrated in the left panel of Fig.\ref{dyonicddq1T0}. In contrast, $\sigma_{xy}(T=0)$ consistently decreases with increasing $\alpha$, as shown in the right panel of Fig.\ref{dyonicddq1T0}). Comparing this behavior with our current model (see the left plot in Fig.\ref{ddqvsk}), we observe that $\sigma_{xx}(T=0)$ behaves differently. Given that $\alpha$ represents the strength of disorder, one would anticipate that conductivities should always be suppressed by this disorder. Therefore, the increase of $\sigma_{xx}$ with $\alpha$ in the dyonic model warrants further investigation.
\begin{figure}[ht]
	\centering
	\includegraphics[width=8cm]{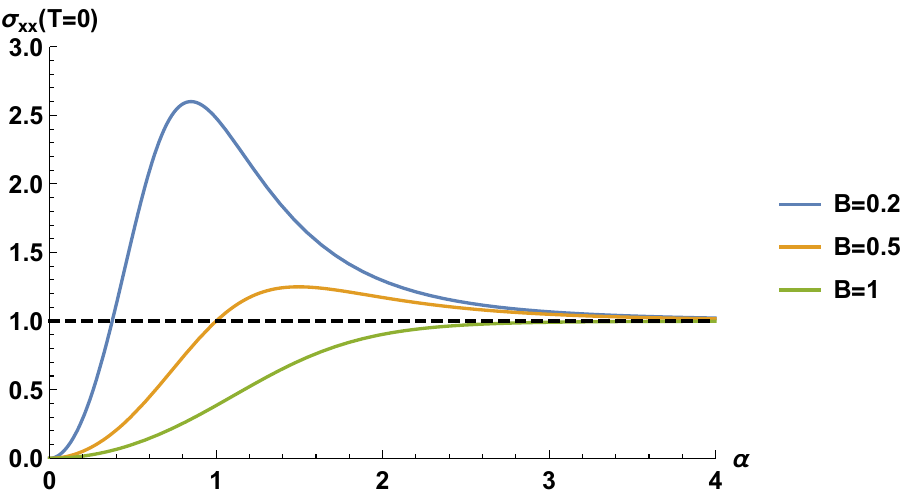}\quad  \includegraphics[width=8cm]{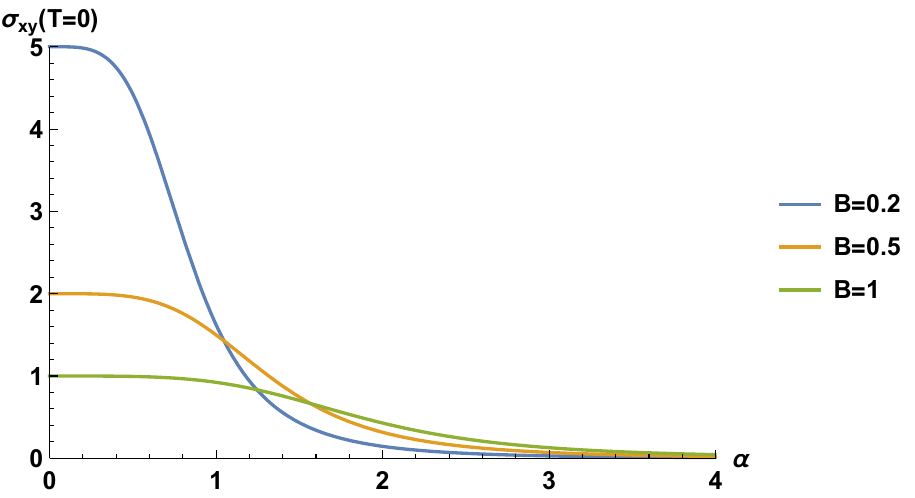}\\
	\caption{$\sigma_{xx}(T=0), \sigma_{xy}(T=0)$ as a function of $\alpha$ in the canonical ensemble.}\label{dyonicddq1T0}
\end{figure}

We are also interested in studying the behavior of $\sigma_{xx}(T=0)$ and $\sigma_{xy}(T=0)$ with respect to the external magnetic field $B$. To achieve this, we will calculate their derivatives with respect to $B$ as follows:
\begin{align}
	&\frac{d\sigma_{xx}(T=0)}{dB}=-\frac{12 B \alpha^2 (12 B^2+\alpha^4+24)}{(12 B^2+\alpha^4)^2 \sqrt{12 B^2+\alpha^4+12}}\\
	&\frac{d\sigma_{xy}(T=0)}{dB}=\frac{12 (\alpha^4-12 B^2)}{(12 B^2+\alpha^4)^2}
\end{align}
It is evident that when $B<\alpha^2/\sqrt{12}$, $\sigma_{xx}$ decreases while $\sigma_{xy}$ increases with increasing $B$. This behavior can be attributed to the deflection of charge carriers caused by the Lorentz force. However, in the regime where $B$ exceeds $\alpha^2/\sqrt{12}$, an intriguing phenomenon emerges: $\sigma_{xy}$ unexpectedly decreases as $B$ increases, mirroring the behavior in our current model. This unexpected phenomenon warrants further investigation and understanding.

We now shift our focus to investigate the temperature-dependent behavior of conductivity, specifically $\sigma_{xx}$, within the dyonic model. To achieve this, we will calculate the derivative of $\sigma_{xx}$ with respect to temperature $T$, as follows:
\begin{align}
	\frac{d\sigma_{xx}}{dT}=\frac{32 \pi  \alpha^2 u_h^3 ((B^3+B)^2 u_h^4+(B^2-1) \alpha^4+2 B^2(B^2+1) \alpha^2 u_h^2)}{(2 B^2 \alpha^2 u_h^2+B^2 (B^2+1) u_h^4+\alpha^4)^2 (3 ((B^2+1) u_h^4+4)+2 \alpha^2 u_h^2)}
\end{align}
To determine whether $d\sigma_{xx}/dT$ is positive or negative, we can equivalently assess the sign of the following quadratic function:
\begin{align}
	Y(u_h)=(B^3+B)^2 u_h^4+(B^2-1) \alpha^4+2 B^2(B^2+1) \alpha^2 u_h^2\,.
\end{align}
The roots of $Y(u_h) = 0$ are readily identified as $u_h^2 = -(B+1)k^2/(B^3+B)$ and $(1-B)k^2/(B^3+B)$. Thus, we can draw the following conclusion:
\begin{align*}
	B<1\qquad \begin{cases} ~(1)~B<1\,,\qquad \alpha^4>\dfrac{12 B^2 (B^2+1)}{1-B^2}   &  \dfrac{d\sigma}{dT}<0  ~~\text{always}\\\\
		~(2)~B<1\,,\qquad \alpha^4<\dfrac{12 B^2 (B^2+1)}{1-B^2} &  \dfrac{d\sigma}{dT}>0\rightarrow \dfrac{d\sigma}{dT}<0 ~~\text{from $T=0$ to high $T$}\\\\
		~(3)~B>1\,,   &  \dfrac{d\sigma}{dT}>0  ~~\text{always}\end{cases}
\end{align*}
The behavior of $\sigma_{xx}(T)$ is also clearly illustrated in Fig.\ref{dyonicddq1TT}.
\begin{figure}[ht]
	\centering
	\includegraphics[width=10cm]{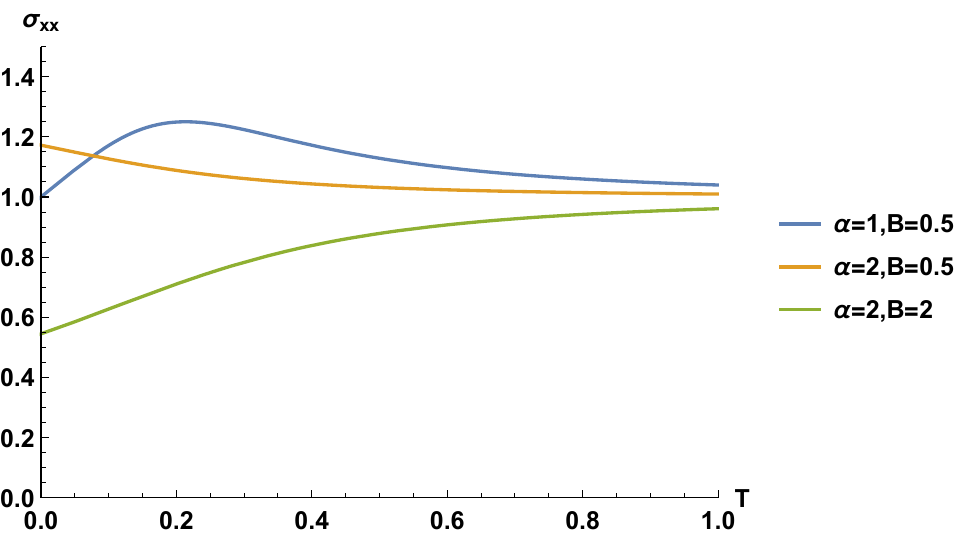}\\
	\caption{$\sigma_{xx}$ as a function of $T$ in the canonical ensemble.}\label{dyonicddq1TT}
\end{figure}
Clearly, we observe that case (1) corresponds to a metallic phase, while case (3) corresponds to an insulating phase. In case (2), as the temperature decreases, the system undergoes a transition from the metallic phase to the insulating phase. These phenomena bear a resemblance to those studied in Section \ref{pro-ec}. This, in turn, suggests that the novel coupling $\mathcal{J}$ indeed plays a role analogous to that of an external magnetic field.

\section{Conclusion and Discussion}

In this work, we investigate the transport properties of a holographic model featuring a novel gauge-axion coupling. As a building block, we introduced a direct coupling between the axion fields with the antisymmetric tensor and the gauge field in our bulk theory.

Interestingly, the presence of the novel coupling term gives rise to nondiagonal components in the conductivity tensor. Furthermore, the diagonal elements exhibit symmetry, while the off-diagonal elements display antisymmetry, i.e., $\sigma_{xx}=\sigma_{yy},\sigma_{xy}=-\sigma_{yx}$, indicating that there should not be anomalous Hall transitions as observed in \cite{Ji:2022ovs}. Notably, we did not introduce a magnetic field into the background, yet the conductivity within this model displays the behaviors akin to Hall conductivity.}

For such a phenomenon, we interpret the result associated with the gauge-axion coupling as the emergence of an ``induced'' magnetic field. We postulate the presence of a magnetic moment within the material, and when subjected to an external electric field, this leads to the spontaneous generation of a magnetic field within the material. We posit that magnetoelectric coupling effects in strongly correlated materials may offer an explanation for this observation. The magnetoelectric coupling effect is a physical phenomenon wherein a magnetic field influences the magnitude of electric polarization or an electric field impacts magnetization. Notably, ferroelectric materials exhibit stable spontaneous polarization and can be manipulated by an applied electric field to generate an internal magnetic field. However, experimental verification of the magnetoelectric coupling effect remains elusive, and we anticipate that future research will provide a more comprehensive understanding and analysis of this phenomenon.

\section*{Acknowledgments}

This work is supported by the Natural Science Foundation of China under Grant No.12375055. J.-P.W. is also supported by the Top Talent Support Program from Yangzhou University.

\appendix
\section{HOLOGRAPHIC RENORMALIZATION}\label{renormalisation}

In this appendix, we present some valuable results essential for carrying out the holographic renormalization procedure. We begin with this process by presenting the formulas for the metric, vector potential, and scalar field in the Fefferman-Graham (FG) coordinate system:
\begin{align}
	&ds^2=\frac{1}{z^2}(dz^2+g_{ij}(z,x)dx^idx^j)\,,\qquad A=A_i(z,x)dx^i\,,\qquad \psi^I=\psi^I(z,x)\,,
\end{align}
and their expansions near the boundary $z=0$:
\begin{subequations}
	\label{FGexp}
	\begin{align}
		&g(z,x)=g_0(x)+g_1(x) z+g_2(x) z^2+g_3(x) z^3+\cdots\,,\\
		&A_i(z,x)=A_{0i}(x)+A_{1i}(x)z+A_{2i}(x)z^2\cdots\,,\\
		&\psi^I(z,x)=\psi^I_0+\psi^I_1(x)z+\psi^I_2(x)z^2\cdots\,.
	\end{align}
\end{subequations}
In addition, in the FG coordinate, the equations of motion are expressed as follows:
\begin{subequations}\label{EMAFGeeq}
	\begin{align}
		&-\mathrm{Tr}(\frac{1}{2}g^{-1}g'')+\frac{1}{4}\mathrm{Tr}(g'g^{-1}g'g^{-1})+\frac{1}{2z}\mathrm{Tr}(g'g^{-1}) =\frac{\Lambda+3}{z^2}+\frac{1}{2}(\psi'_I)^2+\mathcal{O}(z^2) \label{EMAFGeeqzz}\\
		&g^{jk}(-D_ig'_{jk}+D_jg'_{ik})=\psi'_I\partial_i\psi_I+z^2 F_{ij}A'^j\nonumber\\
		&\qquad \qquad\qquad\qquad\qquad\qquad+z^2\mathcal{J}\Big(\epsilon^{IJ}\psi'_I\partial^j\psi_JF_{ij}+\epsilon^{IJ}\partial_i\psi_I\partial^j\psi_JA'_j\Big)+\mathcal{O}(z^3) \label{EMAFGeeqzi}\\
		&R_{ij}[g]-\frac{1}{2}g''_{ij}+\frac{1}{2z}\mathrm{Tr}(g^{-1}g')g_{ij}-\frac{1}{4}\mathrm{Tr}(g^{-1}g')g'_{ij}+\frac{1}{z}g'_{ij}+\frac{1}{2}(g'g^{-1}g')_{ij}\nonumber\\
		&\qquad\qquad  \qquad\qquad \qquad\qquad=\frac{\Lambda+3}{z^2}g_{ij}+\frac{1}{2}\partial_i\psi_I\partial_j\psi_I+\mathcal{O}(z^2)\,,\label{EMAFGeeqii}\\
		&\partial_i\Big(\sqrt{-g}(A'^i+\mathcal{J}\epsilon^{IJ}\psi'^I\partial^i\psi^J)\Big)=0\,,\label{EMAFGmeqzz}\\
		&\partial_z\Big(\sqrt{-g}(A'^i+\mathcal{J}\epsilon^{IJ}\psi'_I\partial^i\psi_J)\Big)+\partial_j\Big(\sqrt{-g}(\tilde{F}^{ji}+\mathcal{J}\epsilon^{IJ}\partial^j\psi_I\partial^i\psi_J)\Big)=0\,,\label{EMAFGmeqzi}\\
		&\partial_z\Big(\frac{\sqrt{-g}}{z^2}(a\psi'^I+z^2\mathcal{J}\epsilon^{IJ}A'^i\partial_i\psi^J)\Big)\nonumber\\
		&\qquad \qquad +\partial_i\Big(\frac{\sqrt{-g}}{z^2}(\partial^i\psi^I-z^2\mathcal{J}\epsilon^{IJ}A'^i\psi'^J+z^2\mathcal{J}\epsilon^{IJ}\tilde{F}^{ij}\partial_j\psi^J)\Big)=0
	\end{align}
\end{subequations}
where the indices are raised and lowered using $g_{ij}$, which is also employed for taking traces. $R_{ij}[g]$ is the Racci tensor of $g_{ij}$, and $\tilde{F}^{ij}\equiv g^{im}g^{jn}F_{mn}$. By examining the equations of motion near the boundary, one can determine the coefficients that satisfy the following relations:
\begin{subequations}
	\label{FGcoeff1}
	\begin{align}
		&\Lambda=-3\,,\qquad g_1=0\,, \qquad \psi_1^I=0\,,\\
		&g_{2ij}=-\Big(R_{ij}^{(0)}[g]-\frac{R^{(0)}[g]}{4} g_{0ij}\Big)+\frac{1}{2}\partial_i \psi^I_0\partial_j \psi^I_0-\frac{1}{8}(\partial_k \psi^I\partial^k \psi^I_0) g_{0ij}\,,\label{d=3EAmodelmetricasyg2}\\
		&\psi^I_2=\frac{1}{2}D^{(0)}_iD^{(0)i}\psi_0^I\,,\label{d=3EAmodelmetricasypsi2}\\
		&\mathrm{Tr}(g_0^{-1}g_3)=0\,,\qquad D^{(0)}_i A^i_1=0\,,\label{widi1}\\
		&D^{(0)j}g_{3ij}=\psi_3^I\partial_i\psi^I_0+\frac{1}{3}F_{0ij}A_{1}^j+\frac{\mathcal{J}}{3}\Big(\epsilon^{IJ}\partial_i\psi_0^I\partial_j\psi_0^JA_{1}^j\Big)\,.\label{widi2}
	\end{align}
\end{subequations}
Here, $g_{0ij}$ is employed for index raising and trace calculation. $R^{(0)}_{ij}[g]$, $R^{(0)}[g]$, and $D^{(0)}_i$ represent the Ricci tensor, curvature scalar, and covariant derivative of $g_{0ij}$, respectively. Additionally, $F_{0ij}=\partial_iA_{0i}-\partial_jA_{0i}$.

With the asymptotic solution at hand, we can readily work out the on-shell renormalized action, which is given by:
\begin{align}
	S_{ren}&=S+\int d^3x \sqrt{-\gamma}(2K-4-R[\gamma]+\frac{1}{2}\gamma^{ij}\partial_i\psi^I\partial_j\psi^I)\,,\label{EMXFXRenAction}
\end{align}
Then, its variation can also be calculated as follows:
\begin{align}
	\delta S_{ren}=&\int d^3x \sqrt{-\gamma}\Big(-K^{ij}+K\gamma^{ij}-2\gamma^{ij}+(R^{ij}[\gamma]-\frac{R[\gamma]}{2}\gamma^{ij})\nonumber \\
	&-\frac{1}{2}\partial^i\psi^I\partial^j\psi^I+\frac{1}{4}(\gamma^{mn}\partial_m\psi^I\partial_n\psi^I)\gamma^{ij}\Big)\delta\gamma_{ij}\nonumber \\
	&-\int d^3x \sqrt{-\gamma}n_\mu ( F^{\mu\nu}+\mathcal{J}X^{\mu\nu})\delta A_\nu\nonumber \\
	&+\int d^3x \sqrt{-\gamma}\Big(-n_\mu(\partial^\mu\psi^I+\mathcal{J}\epsilon^{IJ}F^{\mu\nu}\partial_\nu\psi^J)-\gamma^{ij}D_iD_j\psi^I\Big)\delta\psi^I \label{EMXFXvarRenAction}
\end{align}
The indices here are raised using $\gamma_{ij}$, which represents the components of the induced metric. The symbol $n_\mu$ denotes the outward-pointing unit normal vector of the boundary, while $\gamma$ signifies the determinant of $\gamma_{ij}$, i.e., $\gamma=\det(\gamma_{ij})$. In this context, $R^{(0)}_{ij}[\gamma]$, $R^{(0)}[\gamma]$, and $D_i$ correspond to the Ricci tensor, curvature scalar, and covariant derivative of $\gamma_{ij}$, respectively.
On the other hand, $K^{ij}$ represents the components of the external curvature $K^{\mu\nu}=\gamma^{\mu\alpha}\gamma^{\nu\beta}\nabla_{\alpha}n_\beta$, and $K$ stands for $\nabla_\mu n^\mu$. Furthermore, in the Fefferman-Graham coordinate system, $\delta S_{ren}$ can be expressed as follows:
\begin{align}
	\delta S^{os}_{ren}=\int d^3x \sqrt{-g_0}\Big(\frac{3}{2}(g^{-1}_0g_3g^{-1}_0)^{ij}\delta g_{0ij}+A_{1}^i\delta A_{0i}+(3\psi_3^I+\mathcal{J}\epsilon^{IJ}A_1^i\partial_i\psi^J_0)\delta \psi_{0}^I\Big)\,,\label{FGvarRenAction}
\end{align}
And again, the indices are raised using $g_{0ij}$. Therefore, we can write the one point functions as
\begin{subequations}
	\label{FGTJ}
	\begin{align}
		&\bb T^{ij}\kk=\frac{2}{\sqrt{-g_0}}\frac{\delta S_{ren}}{\delta g_{0ij}}=3(g^{-1}_0g_3g^{-1}_0)^{ij}\,,\\
		&\bb J^i\kk=\frac{1}{\sqrt{-g_0}}\frac{\delta S_{ren}}{\delta A_{0i}}=g_0^{ij}A_{1j}\,,\\
		&\bb O^I\kk=\frac{1}{\sqrt{-g_0}}\frac{\delta S_{ren}}{\delta \psi_{0}^I}=3\psi_3^I+\mathcal{J}\epsilon^{IJ}A_1^i\partial_i\psi^J_0\,.
	\end{align}
\end{subequations}
By virtue of \eqref{widi1} and \eqref{widi2}, we can derive the Ward identities as follows:
\begin{align}
	&\bb T^i_{~~i} \kk=0\,,\qquad   D^{(0)}_i\bb J^i\kk=0\,,\qquad D^{(0)}_j\bb T^{ij}\kk=F^{i}_{0~j}\bb J^j\kk+\bb O^I\kk\partial^i\psi^I_0\,.\label{tlcc}
\end{align}
Note that the last identity can be rewritten as:
\begin{align}
	D^{(0)}_j\bb T^{ij}\kk=(F^{i}_{0~j}+\mathcal{J}\epsilon^{IJ}\partial^i\psi^I_0\partial_j\psi^J_0)\bb J^j\kk+3\psi_3^I\partial^i\psi^I_0\,,\label{tlcc}
\end{align}
It implies that time reversal symmetry can be broken even without the presence of an external magnetic field. For instance, consider the case where $F^{x}_{0~y}=0$. In the evolution of the momentum density $T^{xt}$, there are two couplings to consider: $F^{x}_{0~t}\bb J^t\kk$ and $\mathcal{J}\epsilon^{IJ}\partial^x\psi^I_0\partial_j\psi^J_0\bb J^j\kk$. The first coupling involves only the charge density $\bb J^t\kk$, thereby preserving time reversal symmetry, whereas the second coupling includes the charge current density $\bb J^x\kk$ and $\bb J^y\kk$, potentially leading to a breaking of time reversal symmetry. This is the reason we observe $\sigma_{xy}=-\sigma_{yx}$ even in the absence of an external magnetic field. We will elucidate this point in what follows.

As we know, the off-diagonal elements of conductivity satisfies $\sigma_{xy}=\sigma_{yx}$ under time reversal symmetry. This comes from the property of the response function
	\begin{align}\label{TT}
	\chi_{ij}(\omega)=\varepsilon_i\varepsilon_j\chi_{ji}(\omega)\,.
	\end{align}
where $\varepsilon_i=\pm1$ is signature of the operater under time reversal \cite{Hartnoll:2009sz}. In our case, time reversal symmtry is  broken. A new symmetry that combining time reversal and operation from $\mathcal{J}$ to $-\mathcal{J}$ are established. Consequently, \eqref{TT} is replaced by $\chi_{ij}(\omega,\mathcal{J})=\varepsilon_i\varepsilon_j\chi_{ji}(\omega, -\mathcal{J})$, and we have
	\begin{align}\label{TTJ}
	\sigma_{xy}(\omega,\mathcal{J})=\sigma_{yx}(\omega,-\mathcal{J})\,.
    \end{align}
This is similar to the result of the magnetic field leading to the breaking of the time reversal invariant, i.e.,  $\chi_{ij}(\omega,B)=\varepsilon_i\varepsilon_j\chi_{ji}(\omega,-B)$. $\sigma_{xy}$ is an odd function of $\mathcal{J}$ (see Eq.\eqref{xy}), so Eq.\eqref{TTJ} is visible obviously. Therefore, we conclude that the breaking of time reversal symmetry is responsible for the emergence of the antisymmetric conductivity in our model.

	\bibliographystyle{style1}
	\bibliography{Ref}
\end{document}